\documentstyle[12pt,epsfig]{article}

\begin{document}

\parindent 1.3cm
\thispagestyle{empty} 
\vspace*{-3cm}
\noindent

\def\gt{{\tilde g}}
\def\ttb{(T,\overline T)}
\def\sa{{\rm s}_\alpha}
\def\ca{{\rm c}_\alpha}
\def\m12t{\tilde m_{12}^2}
\def\BR{\rm BR}
\def\arccot{\mathop{\rm arccot}\nolimits}
\def\sd{\strut\displaystyle}

\begin{obeylines}
\begin{flushright}
\vspace{1cm}
UG-FT-92/98
\end{flushright}
\end{obeylines}
\vspace{3cm}

\begin{center}
\begin{bf}
\centerline {THE LIGHT HIGGS IN SUPERSYMMETRIC MODELS}
\centerline {WITH HIGGS TRIPLETS}
\end{bf}
\vspace{1.5cm}

M. Masip 

\vspace{0.1cm}
{\it Departamento de F\'\i sica Te\'orica y del Cosmos\\
Universidad de Granada\\
18071 Granada, Spain\\}
\vspace{2.2cm}

{\bf ABSTRACT}

\parbox[t]{12cm}{
In supersymmetric models 
the presence of Higgs triplets introduce new quartic interactions
for the doublets that may raise the mass of the lightest 
$CP$-even field up to 205 GeV. We show that 
the complete effect of the triplets can 
be understood by decoupling them from the minimal sector and then 
analyzing the vacuum and the spectrum of the effective two-Higgs 
doublet model that results. 
We find that the maximum value of $m_h$ is only achieved in a
very definite region of the parameter space. In this region,
however, radiative corrections 
decrease the bound to $\approx 190$ GeV.
}\end{center}
\vspace*{3cm}

\newpage

\noindent {\bf 1. Introduction.} 
In the minimal supersymmetric standard model 
(MSSM) the quartic couplings of the Higgs fields are not
free parameters, they are related to the $SU(2)_L\times U(1)_Y$
gauge couplings \cite{hhg90}. This implies that the tree-level 
mass of the lightest $CP$-even scalar field is smaller than $M_Z$:
\begin{equation}
m_h^2\le M_Z^2 \cos^2 2\beta\;,
\label{bound0}
\end{equation}
where $\tan\beta$ is the ratio $\langle H_2\rangle/\langle 
H_1\rangle$ of vacuum expectation values (VEVs) of the two Higgs 
doublets.

There are two different mechanisms that can raise this upper bound
on $m^2_h$. The first one is due to the fact that supersymmetry 
(SUSY) is broken. The SUSY values of the quartic couplings suffer 
radiative corrections \cite{hab93} from top loops 
proportional to $h_t^4\ln (m_{\tilde t}^2/m_t^2)$, where $h_t$ is 
the top Yukawa coupling and $m_t$ and $m_{\tilde t}$ are the 
fermion and the scalar masses, respectively. This translates 
into corrections to the Higgs mass of order 
$\Delta m_h^2\approx (90$ GeV$)^2$. Also the soft SUSY-breaking 
trilinear $V\supset A_t\; \tilde t^c\tilde t H^0_2$ may have an
impact on $m_h^2$. At one loop it gives nonlogarithmic corrections 
\cite{oka91} proportional to $h_t^2 {A_t^2\over m_{\tilde t}^2}-
{1\over 12}{A_t^4\over m_{\tilde t}^4}$ that can be of the same size.

The other mechanism to increase the bound on $m_h^2$ requires the 
presence of extra Higgs fields and couplings, namely, gauge 
singlets \cite{kin96} or triplets \cite{esp98} and couplings of 
these fields with the doublets. Trilinears in the superpotential 
$W$ involving two doublets and the extra field introduce new 
quartic interactions for the scalar doublets and then new 
contributions to $m_h^2$.

When introducing extra Higgs fields, however, there is another 
effect {\it competing} with the positive effect on $m_h^2$ of the
quartic couplings. It is due to the mixing between the lightest
and heavier states. Such a mixing tends to decrease 
the smallest eigenvalue in the Higgs mass matrix. In a particular
escenario one would expect that the new (arbitrary) parameters 
present, soft masses and trilinears, may be adjusted in order to
cancel the mixing. In that case, for real VEVs the lightest neutral 
state would be obtained diagonalizing the $2\times 2$ submatrix 
defined by the $CP$-even scalars in the doublets. Then the maximum
value of $m_h$ would be obtained for heavy $CP$-odd states.

In particular, let us consider the presence of a pair of $SU(2)_L$ 
triplets $\ttb$ of $(-1,+1)$ hypercharge:
\begin{equation}
T=\left( \begin{array}{c}
T^0 \\ T^- \\ T^{--}
\end{array} \right)\; ; \;\; 
\overline T = \left( \begin{array}{c}
\overline T^{++} \\ \overline T^+ \\ \overline T^0
\end{array} \right) \;. 
\label{phi}
\end{equation}
These fields admit a term in the superpotential of type
\begin{equation}
W\supset {1\over 2}\chi\; ( T^0 H_2^0 H_2^0 - 
\sqrt{2}\; T^- H_2^0 H_2^+ + T^{--} H_2^+ H_2^+)\;.
\label{w}
\end{equation}
An analogous term can be obtained exchanging 
$T \leftrightarrow \overline T$, $H_2\leftrightarrow H_1$.
With the quartic coupling 
$V\supset {\chi^2\over 4}\; |H^0_2 H^0_2|^2$ 
in the scalar potential the bound on $m_h^2$  becomes
\begin{equation}
m_h^2\le M_Z^2 \cos^2 2\beta + \chi^2 v^2 \sin^4\beta\;,
\label{bound1}
\end{equation}
where $v^2=\langle H_1\rangle^2+\langle H_2\rangle^2\approx (174\;
{\rm GeV})^2$. Although $\chi$ is in principle a free parameter,  
it is constrained by the following argument. Its evolution with 
the energy scale is given by
\begin{equation}
{{\rm d} \chi \over {\rm d}t}={\chi\over 16 \pi^2}\;
(\;{7\over 2} \chi^2 + 6 h_t^2 - {9\over 5} g_1^2 - 7 g_2^2\;)\;.
\label{rge}
\end{equation}
The possibility to integrate 
the electroweak and the grand unification (GUT) scales is here
a main motivation, as in any SUSY model. However, large initial 
values of $\chi$ would become nonperturbative before the GUT 
scale $M_X$. Evolving the model down from $M_X$ one finds that 
at low energies $\chi$ is always smaller than $\approx 0.9$. This 
value could be enough to put the bound in Eq.~(\ref{bound1}) 
around 205 GeV \cite{esp98}. 

Our objective in this letter is to find out whether the bound 
in Eq.~(\ref{bound1}) can be saturated or not. We discuss a 
simple way to understand the effect of adding extra triplets 
(or any other vectorlike Higgs field) of mass 
$m= M+O(m_{SUSY})$. We decouple these scalar fields from the 
minimal sector keeping the terms of first 
order in $1/m$. The effective model that results is a particular 
two-Higgs doublet model that depends on the physics of the 
triplets. The decoupling effects are equivalent to the
mixing in the complete mass matrix, but much simpler to analyze. 
The method is justified when 
$m^2\gg (v^2,{\chi^2\over 4\pi}m_{SUSY}^2)$.
In the triplet model this seems to be a necesary requirement
because the VEVs grow as the inverse of their mass, 
and if $\langle T,\overline T \rangle \ge 10$ GeV 
it predicts an unacceptable value of the $\rho$ 
parameter (a different Weinberg angle measured 
from gauge boson masses and charged currents) \cite{lan91}. 
We compare the value for $m_h$ obtained in the 
effective model with the exact numerical solution in the 
complete triplet model and find that the agreement is good
even for $m\approx m_{SUSY}$. The method allows us to
identify the only region in the parameter space that saturates
the bound in Eq.~(\ref{bound1}). This region has a very definite
pattern of SUSY breaking terms, with large radiative corrections
that partially cancel top quark effects. 

\noindent {\bf 2. The triplet model.} 
Let us consider the neutral Higgs sector of the model
(from now on we drop the $0$ superscript to indicate 
neutral fields). We include in $W$ the terms
\begin{equation}
W\supset-\mu\; H_1H_2-M\; \overline T T + 
{\chi\over 2}\; T H_2 H_2\;.
\label{wt}
\end{equation}
Adding soft SUSY-breaking terms the relevant part of the 
scalar potential is 
\begin{eqnarray}
V&=&m_1^2\; H_1^\dagger H_1 + m_2^2\; H_2^\dagger H_2 + 
m_3^2\; T^\dagger T + m_4^2\; \overline T^\dagger 
\overline T \nonumber \\
&&
- (m_{12}^2\; H_1 H_2 + {\rm h.c.})
+(A_M^2 \; \overline T T + {\rm h.c.})-
(A_\chi \; T H_2 H_2 + {\rm h.c.})
\nonumber \\
&&
-({1\over 2}\chi M\;\overline T^\dagger H_2 H_2+{\rm h.c.})
-(\chi\mu \; H_1^\dagger T H_2 + {\rm h.c.}) +
\chi^2\; H_2^\dagger T^\dagger H_2 T
\nonumber \\
&&
+{1\over 4} \chi^2\; (H_2^\dagger H_2)^2
+\gt\; (H_1^\dagger H_1-H_2^\dagger H_2+
2 T^\dagger T-2\overline T^\dagger\overline T)^2
\;,\nonumber \\
\label{vt}
\end{eqnarray}
where $\gt=(g_Y^2+g_L^2)/8$ and all the fields are neutral. 
Field redefinitions can be used to
set $m_{12}$, $A_\chi$ and $\chi M$ real and positive. We 
asume for simplicity 
that $A_M^2$ and $\chi \mu$ are real, but this does not guarantee
that all the VEVs are real and positive (which would be the case if 
$\chi\mu\ge 0$ and $A_M^2\le 0$).
The size of the mass parameters above is
$(m_1^2,m_2^2,m_{12}^2) = O(m_{SUSY}^2)$; 
$(\mu, A_\chi) =  O(m_{SUSY})$; 
$M \ge  O(m_{SUSY})$; 
$(m_3^2,m_4^2) =M^2+O(m_{SUSY}^2)$; and $A_M^2=O(m_{SUSY}M)$.

It is convenient to express the fields in terms of 
moduli and phases:
\begin{eqnarray} 
H_1={1\over \sqrt{2}} v_1 e^{\theta_1} &;&\;\;
H_2={1\over \sqrt{2}} v_2 e^{\theta_2}\; ;\nonumber \\
T={1\over \sqrt{2}} v_3 e^{\theta_3} &;&\;\;\;\;
\overline T={1\over \sqrt{2}} v_4 e^{\theta_4}\; .\nonumber \\
\label{vevs4}
\end{eqnarray}
The determination of the minimum of the potential in 
Eq.~(\ref{vt}) and of $m_h^2$ requires much algebra (in the
general case with complex VEVs, the diagonalization of a 
$8\times 8$ matrix). 
We propose, instead, to integrate the fields 
$\ttb$ out, analize the effective model that results, and check 
numerically that
for any particular choice of parameters the complete and
the approximate models give the same spectrum for the four 
lightest fields.

\noindent {\bf 3. The approximate model.}
We first rewrite $\ttb$ in terms of mass eigenstates:
\begin{equation}
T_1=\ca T - \sa \overline T^\dagger \; ;\;\;
T_2=\sa T + \ca \overline T^\dagger \;,
\label{eigenstates}
\end{equation}
where $\sa = \sin\alpha$, $\ca = \cos\alpha$ and
$\tan 2\alpha=(2A_M^2)/(m_4^2-m_3^2)$. Their masses are
\begin{equation}
M_1^2=\ca^2 m_3^2 +\sa^2 m_4^2 - 2\sa\ca A_M^2\; ;\;\;
M_2^2=\sa^2 m_3^2 +\ca^2 m_4^2 + 2\sa\ca A_M^2\;.
\label{eigenvalues}
\end{equation}
If $M\gg m_{SUSY}$ then $\alpha\approx {\pi\over 4}$, 
$M_1^2\approx M^2-A_M^2$ and $M_2^2\approx M^2+A_M^2$.

Integrating $T_1$ and $T_2$ out it results the 
two Higgs doublet model
\begin{eqnarray}
V&=&m_1^2\; H_1^\dagger H_1 + m_2^2\; H_2^\dagger H_2 
- (m_{12}^2\; H_1 H_2 + {\rm h.c.})\nonumber \\
&&
+\lambda_3\; H_1^\dagger H_1 H_2^\dagger H_2
+({1\over 2}\lambda_7\; H_2^\dagger H_1 H_2 H_2 + {\rm h.c.})
+{1\over 4} \tilde\chi^2\; (H_2^\dagger H_2)^2
\nonumber \\
&&
+\gt\; (H_1^\dagger H_1-H_2^\dagger H_2)^2
\;,\nonumber \\
\label{vapp}
\end{eqnarray}
where the contributions of the triplets to 
$\tilde \chi^2$, $\lambda_3$ and $\lambda_7$ come 
from the diagrams (a), (b) and (c) in Fig.~1, 
respectively. We obtain 
\begin{eqnarray} 
\tilde \chi^2&=&\chi^2-
{(\chi M \sa - 2 A_\chi\ca)^2\over M_1^2}-
{(\chi M \ca + 2 A_\chi\sa)^2\over M_2^2}\;,
\nonumber \\
\lambda_3&=&-{(\chi \mu \ca)^2\over M_1^2}-
{(\chi \mu \sa)^2\over M_2^2}\;,\nonumber \\
\lambda_7&=&-{(\chi \mu \ca)
(2 A_\chi\ca-\chi M \sa)\over M_1^2}-{(\chi \mu \sa)
(2 A_\chi\sa + \chi M \ca)\over M_2^2}\;.\nonumber \\
\label{couplings}
\end{eqnarray}
Higher dimensional operators would introduce corrections
of order $v^2/M_{1,2}^2$, whereas one-loop corrections will 
be small if ${\chi^2\over 4 \pi}m_{SUSY}^2 < M_{1,2}^2$.

To find the minimum and the spectrum of this model 
we express the fields in terms of moduli and phases:
\begin{equation} 
H_1={1\over \sqrt{2}} v_1 e^{\theta_1}\; ;\;\;
H_2={1\over \sqrt{2}} v_2 e^{\theta_2}\; .
\label{vevs}
\end{equation}
Then
\begin{eqnarray}
V&=&{1\over 2} m_1^2\; v_1^2 + {1\over 2} m_2^2\; v_2^2 
- m_{12}^2\; v_1 v_2 \cos (\theta_1+\theta_2)\nonumber \\
&&
+{1\over 4}\lambda_3\; v_1^2 v_2^2
+{1\over 4}\lambda_7\; v_1 v_2^3 \cos (\theta_1+\theta_2)
+{1\over 16} \tilde\chi^2\; v_2^4
\nonumber \\
&&
+{\gt\over 4}\; (v_1^2 - v_2^2)^2
\;.\nonumber \\
\label{vapp2}
\end{eqnarray}
We can use an hypercharge transformation to set 
$\langle \theta_1\rangle =0$. 
The minimum conditions give then $\langle \theta_2\rangle=0$,
$\langle v_1\rangle$ and $\langle v_2\rangle$. 
In the $4\times 4$ mass matrix the $CP$-odd sector, 
$M_{ij}={1\over v_i v_j}{\partial^2 V\over \partial\theta_i
\partial\theta_j}$, does not mix with the 
the $CP$-even sector, $M_{2\!+\!i\;2\!+\!j}=
{\partial^2 V\over \partial v_i \partial v_j}$. 
We find
\begin{eqnarray}
&&M_{11}=\m12t\tan\beta\;,\nonumber \\
&&M_{12}=\m12t\;,\nonumber \\
&&M_{22}=\m12t\tan^{-1}\beta\;;\nonumber \\
&&M_{33}=\m12t\tan\beta+M_Z^2 \cos^2\beta\;,\nonumber \\
&&M_{34}=-\m12t-M_Z^2 \sin\beta\;\cos\beta 
+ 2\lambda_3\sin\beta\;
\cos\beta+\lambda_7 v^2 \sin^2\beta\;,\nonumber \\
&&M_{44}=\m12t\tan^{-1}\beta+M_Z^2 \sin^2\beta
+\tilde\chi^2 v^2 \sin^2\beta + {3\over 2} 
\lambda_7 v^2\sin\beta\;\cos\beta \;;\;.\nonumber \\
\label{mm}
\end{eqnarray}
where $\m12t=m_{12}^2-{\lambda_7\over 4} \langle v_2\rangle^2$, 
$v^2={\langle v_1\rangle^2+\langle v_2\rangle^2\over 2}$, 
$\tan\beta={\langle v_2\rangle\over \langle v_1\rangle}$ and
$M_Z^2=4\gt v^2$ (we neglect the contribution of
triplet VEVs to $M_Z$).

It is now straightforward to find the mass eigenvalues.
In the $CP$-odd sector there is, in addition to the massless
Goldstone, a field of mass $m_A^2=\m12t/(\sin\beta\cos\beta)$.
The lightest Higgs is in the $CP$-even sector, together with
a field of mass $m_H^2 = O(\m12t)$. $m_h^2$ is bounded to be
smaller than $(M_{33}M_{44}-M_{34}^2)/(M_{33}+M_{44})$, value
that is saturated in the limit $\m12t \gg v^2$. In this
limit we can obtain an approximate expression for $m_h^2$:
\begin{equation}
m_h^2\approx M_Z^2 \cos^2 2\beta + \tilde\chi^2 v^2 \sin^4\beta
+4\lambda_3 v^2 \sin^2\beta \cos^2\beta+
{7\over 2} \lambda_7 v^2 \sin^3\beta \cos\beta
\;,
\label{bound2}
\end{equation}
where the couplings have the value 
specified in Eq.~(\ref{couplings}).

Before discussing how to tune the parameters in order to 
approach the bound in Eq.~(\ref{bound1}), let us check the
efficiency of our approach. We have computed numerically the
spectrum of light fields in the effective and in the complete 
triplet models for many different values of the parameters, 
changing the signs of $A_M^2$ and $\chi\mu$. 
We obtain that the results for $m_h$ in both models 
always agree within a $2\%$ margin. For example, 
let us take a triplet model with $M=3$ TeV, $\chi=0.7$, 
$A_M^2=(0.5{\rm\; TeV}) M$, $A_\chi=0.5 \chi$, $\mu=0.4$ TeV,
$m_{12}^2=0.25{\rm\; TeV}^2$, $m_{3}^2=9.4{\rm\; TeV}^2$, 
$m_{4}^2=9.5{\rm\; TeV}^2$, 
and the mass parameters $m_1^2=0.50{\rm\; TeV}^2$ and
$m_2^2=0.12{\rm\; TeV}^2$ (chosen to have $\tan\beta=2$
and the right value of $M_Z$). The spectrum of this model is
(3310, 2819, 790.7) GeV in the $CP$-odd sector and 
(3310, 2819, 794.3, 51.4) GeV in the $CP$-even sector. The
value in the MSSM that corresponds to this value of $\tan\beta$
is $m_h=55.2$ GeV, versus $m_h=51.4$ GeV obtained here
(both values would coincide if the 
triplets were completely decoupled). Defining an analogous
model with opposite sign for $\chi\mu$ ({\it i.e.}, 
$\mu=-0.4$ TeV) we obtain $m_h=59.3$ GeV. 

The corresponding two-Higgs doublet model is built with
the couplings in  Eq.~(\ref{couplings}): $\tilde \chi=0.10$,
$\lambda_3=-0.0086$, $\lambda_7=-0.0113$, 
and the masses 
$m_1^2=0.50{\rm\; TeV}^2$, $m_2^2=0.12{\rm\; TeV}^2$ 
In order to keep the same value of $M_Z$ and $\tan\beta=2$,
these masses are not identical 
to $m_{1,2}^2$ in the complete triplet model 
(however, the difference is in next digits).
Here we obtain a field of
(790.8) GeV in the $CP$-odd sector and fields of
(794.4, 51.2) GeV in the $CP$-even sector. The
value obtained for the mass of the lightest Higgs, 
$m_h=51.2$ GeV, is very close to the value
$m_h=51.4$ GeV of the complete triplet model. In the 
approximate model that corresponds to $\mu=-0.4$ TeV we obtain
$m_h=59.2$ GeV, also in agreement with the value $m_h=59.3$ GeV 
of the complete model. 
For this choice of $\chi$ and $\tan \beta$
the bound in Eq.~(\ref{bound1}) is $m_h\le 111.9$ GeV, a number
that gives no information.

We also obtain an excellent approximation
when the triplet fields to integrate are not heavier than the
doublets. For example, taking
$m_1^2=1.12{\rm\; TeV}^2$, 
$m_2^2=0.05{\rm\; TeV}^2$, $m_3^2=0.95{\rm\; TeV}^2$, 
$m_4^2=1.00{\rm\; TeV}^2$, 
$M=0.7$ TeV, $\chi=0.7$, 
$A_M^2=(0.5{\rm\; TeV}) M$, $A_\chi=0.5 \chi$, $\mu=-0.4$ TeV,
and $m_{12}^2=0.25{\rm\; TeV}^2$ 
we have $\tan\beta=5$, 
$CP$-odd scalars of (1175, 1119, 791) GeV and 
$CP$-even scalars of (1175, 1119, 791, 91.9) GeV.
Actually, those values correspond to a local minimum in an 
unbounded potential; this is the tendency (due to the large 
number of complex phases) in most of the
parameter space when $M\le m_{SUSY}$. In this case 
the triplets to integrate out are lighter 
than one of the doublets. In the approximate model
we have $\tilde \chi=-0.296$,$\lambda_3=-0.095$ and 
$\lambda_7=0.18$. We obtain $m_h= 91.2$ GeV, in 
agreement with the result $m_h= 91.9$ GeV in the 
complete model (versus $m_h= 84.9$ GeV in the MSSM or 
$m_h\le 144.7$ GeV in Eq.~(\ref{bound1})).
The numerical analysis shows that the effective two-Higgs
doublet model describes very efficiently the effect of the 
triplets on $m_h$.

\noindent {\bf 4. Maximum value of $m_h$.} 

Now, from the expressions in 
Eqs.~(\ref{couplings},\ref{bound2}) it 
is clear that for a SUSY triplet mass 
$M\ge m_{SUSY}$ the bound in Eq.~(\ref{bound1}) 
is never approached. For $M$ much 
larger than the SUSY-breaking masses $\tilde \chi^2$ 
goes to zero, and the triplets decouple (as expected).
If $2A_\chi\ca$ is large and tends to
cancel $\chi M\sa$ in one of the terms defining 
$\tilde \chi^2$, then in the other 
term $\chi M \ca + 2 A_\chi\sa$ will be large
and $\tilde \chi^2$ goes also to zero.
The contribution to $m_h^2$ proportional to $\lambda_3$
is always negative, whereas the one proportional to
$\lambda_7$ can be positive if $\chi\mu<0$.
However, in this case a sizeable contribution
would require that all the couplings ($\chi M$, $2A_\chi$
and  $\chi\mu$)
are of the same order, implying complex VEVs and 
mixing of the light Higgs with the (heavy) $CP$-odd sector.
Such a mixing also would lower $m_h^2$.
In any case, the $\lambda_{3,7}$ terms are not relevant 
in the region of large $\tan\beta$, where the bound in 
Eq.~(\ref{bound1}), if saturated, allows a light 
Higgs of up to 205 GeV \cite{esp98}.

The maximum value of $m_h$ would be obtained for 
a SUSY mass $M$ and a SUSY-breaking 
trilinear $A_\chi$ both much smaller than the SUSY-breaking 
masses of the triplets: $M^2\ll (M^2_1,M^2_2)$, 
$2A_\chi^2\ll (\chi^2M^2_1,\chi^2M^2_2)$. In this limit
the scalar triplets decouple but the quartic coupling that
they introduce in the Higgs doublet sector remains. In
consequence, the tree-level bound 
in Eq.~(\ref{bound1}) would be approached.

This very definite pattern of SUSY breaking parameters,
however, has obvious implications at the quantum level. 
We have here a large splitting between the fermion and
the scalar components of the triplet superfields and
also a large Yukawa coupling (see Eq.~(\ref{wt})), very
much like in the top quark sector. To estimate the 
radiative effects 
let us focus on the region of large $\tan\beta$, where the
light neutral Higgs $\phi$ is basically $H_2$.
We simplify and assume that 
$M\approx m_t$ and all the SUSY-breaking masses coincide,
$m_{SUSY}\approx 1$ TeV, with a generic suppression of 
scalar trilinears that makes negligible 
nonlogarithmic corrections.
Below $m_{SUSY}$ we have the standard model, with
$V=m^2_\phi\; \phi^\dagger \phi + {\lambda\over 4} \; 
(\phi^\dagger \phi)^2$, 
plus the fermion components of the triplets. 
At $m_{SUSY}$ the quartic scalar coupling is 
\begin{equation}
\lambda_0 = {g_Y^2+g_L^2\over 2}+\chi^2
\;,
\label{lambda}
\end{equation}
and its running down to the electroweak scale is given by
\begin{equation}
{{\rm d} \lambda \over {\rm d}t}={3\over 16 \pi^2}\;
[\;\lambda^2 + (2 h_t^2+\chi^2)\lambda -4h_t^4 
-{5\over 3} \chi^4\;]\;.
\label{rge-lambda}
\end{equation}
In the MSSM one obtains 
$\lambda(m_t^2)\approx \lambda_0+ {3\over 4\pi^2}
h_t^4\; \ln {m_{SUSY}^2\over m_t^2}$ and 
$m_h^2\approx (92^2+ 88^2)$ GeV$^2$. Here there is a partial
cancellation in the $\beta$ function, with 
\begin{equation}
\lambda(m_t^2)\approx \lambda_0+
{3\over 16\pi^2}\; (4 h_t^4 -2h_t^2\chi^2-{1\over 3}
\chi^4)\;\ln {m_{SUSY}^2\over m_t^2}
\label{deltalambda}
\end{equation}
and 
\begin{eqnarray}
m_h^2 &\approx & M_Z^2 + \chi^2 v^2 +
{3\over 16 \pi^2} (
4 h_t^4 - 2 h_t^2\chi^2 - {1\over 3}
\chi^4)v^2\; \ln {m_{SUSY}^2\over m_t^2}\nonumber \\
&\approx & (92^2+156^2+62^2)\;{\rm GeV}^2 = (190
\;{\rm GeV})^2,\nonumber \\
\label{deltamh}
\end{eqnarray}
where we have neglected the electroweak gauge couplings and
the evolution of $h_t$ and $\lambda$
with the scale 
(both with decreasing effect on the
estimated size of radiative corrections).

\noindent {\bf 4. Conclusions.} 
The mass of the Higgs in the MSSM is constrained to be smaller
than $M_Z$ at the tree level and smaller than around
$130$ GeV once 
radiative (SUSY-breaking) corrections are included. In more
general models there are new fields and new 
quartic Higgs interactions raising
$m_h$. However, the new fields also introduce 
mixing with the Higgs doublets, which decreases $m_h$. 
In models with gauge singlets it is easy to see that 
this mixing can be fine tuned to
zero. However, in models with triplets the larger number of 
fields and parameters makes the analysis too complicated. 

We have presented a method that allows to understand
the complete effect of the triplets on $m_h$.
It is based on an effective model that results 
integrating the triplets out and keeping 
only their effect on the quartic couplings of the doublets. 
In this effective model the bound on $m_h$ 
is very simple
to estimate (see Eq.~(\ref{bound2})). 
The procedure gives an excellent approximation for 
triplet masses larger than the electroweak scale $v\approx
174$ GeV, even if the triplets and the doublets have similar
masses.

Using this method we show that, in order to modify
substantially the MSSM values of $m_h$, the scalar triplets 
must have a small SUSY mass (their mass 
must be basically a SUSY-breaking term) and their 
trilinears must be suppressed respect the masses.
The large splitting in the triplet
supermultiplet, with heavy scalars and light fermions,
together with the large Yukawa coupling 
$\chi$ and the absence of scalar trilinears, define a 
very clear pattern of SUSY-breaking parameters with
implications on the size of radiative corrections. We 
have estimated 
these corrections and obtained that they are significantly
smaller than in the MSSM. 

We conclude that the triplet model is able to provide 
values of $m_h$ of up to 190 GeV. The value 205 GeV seems 
an overestimate, since it would require $M\approx m_{SUSY}$ 
and large SUSY-breaking trilinears (it is based on a 
cancellation between trilinears of order $m_{SUSY}$ that we 
show cannot take place). The value $m_h=190$ GeV 
is still larger than $m_h= 155$ GeV \cite{mas98}
of the singlet model with intermediate vectorlike 
matter. Probably, the triplet model 
has ingredients that make it 
a less appealing framework from a model building point of view: 
the need to avoid triplet VEVs, 
the need for four pairs of colour triplets at low energy to 
obtain gauge unification, 
the need to avoid extra matter with electroweak charges at 
intermediate scales ($g_Y$ and $g_L$ are near their perturbative 
fixed point values), or the need to incorporate a generalized
R-parity to be realistic (the usual $Z_2$ matter parity of the 
MSSM cannot avoid here, for example, unacceptable neutrino masses).
However, the pattern of SUSY-breaking terms 
(large scalar masses versus trilinears) and the
presence of double charged leptons at $\approx 200$ GeV
required to saturate the bound define an interesting region 
of its parameter space.
For example, there the triplet VEVs are naturally small:
\begin{equation}
\langle v_4\rangle\approx {\sqrt{2}\chi M v^2 \sin^2\beta \over 
m_4^2}\;,
\label{v4}
\end{equation}
easily within the experimental limit of 10 GeV.

\noindent {\bf Acknowledgments.} 

The author thanks F. del Aguila, R. Mu\~noz-Tapia and
A. Pomarol for discussions.
This work was supported by CICYT under contract 
AEN96-1672 and by the Junta de Andaluc\'\i a under contract
FQM-101.

\newpage
\setlength{\unitlength}{1cm}
\begin{figure}[htb]
\begin{picture}(10,15)
\epsfxsize=5.cm
\put(5.,0.5){\epsfbox{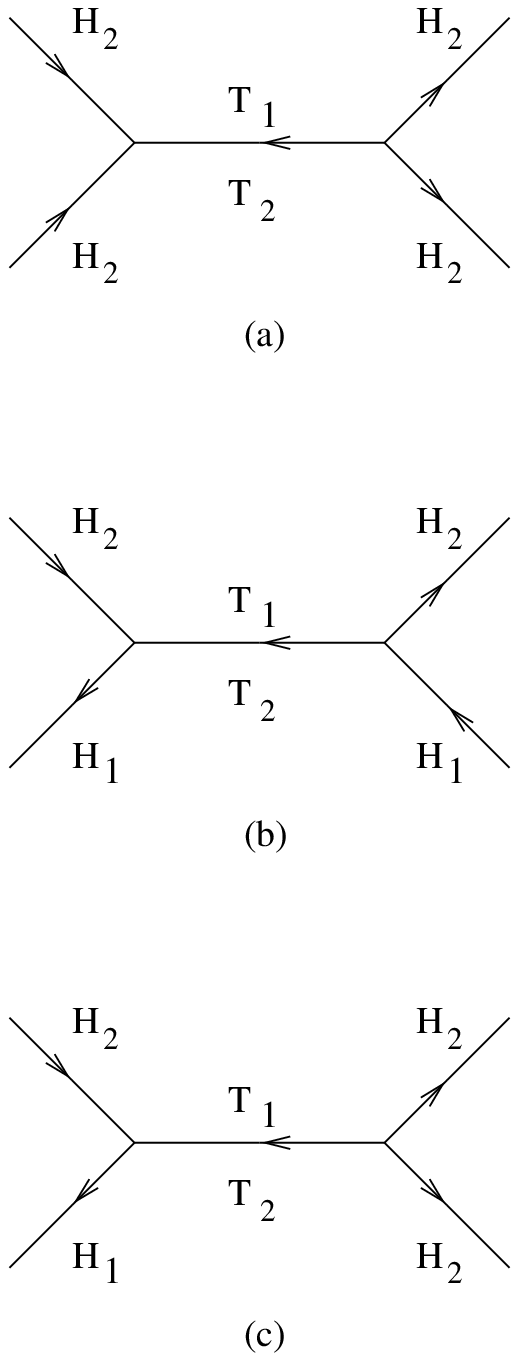}}
\end{picture}
\caption{
Diagrams contributing to $\tilde \chi^2$ (a),
$\lambda_3$ (b) and $\lambda_7$ (c).
\label{fig}}
\end{figure}


\begin{thebibliography}{99}

\bibitem{hhg90} J.F. Gunion, H.E. Haber, G. Kane and
S. Dawson, {\it The Higgs Hunter's Guide}
(Addison-Wesley, Reading, MA, 1990).

\bibitem{hab93} H.E. Haber and R. Hempfling,  
Phys. Rev. D {\bf 48} (1993) 4280.

\bibitem{oka91} Y. Okada, M. Yamaguchi and T. Yanagida,
Phys. Lett. B {\bf 262} (1991) 54.

\bibitem{kin96} S.F. King and P.L. White, 
Phys. Rev. D {\bf 53} (1996) 4049.

\bibitem{esp98} J.R. Espinosa and M. Quir\'os,
Phys. Rev. Lett. {\bf 81} (1995) 516.

\bibitem{lan91} P. Langacker and M. Luo, 
Phys. Rev. D {\bf 44} (1991) 817.

\bibitem{mas98} M. Masip, R. Mu\~noz-Tapia and
A. Pomarol, Phys. Rev. D {\bf 57} (1998) 5340.

\end{thebibliography}
\end{document}